\documentclass[12pt]{article}

\usepackage{graphicx,amssymb,amsmath,tensor,empheq,color}

\usepackage{physics}

\usepackage{hyperref}
\hypersetup{colorlinks = true, linkcolor=black, citecolor=black, urlcolor=blue}

\newlength{\fighskip} \fighskip=2pt
\newlength{\figvskip} \figvskip=3pt

\newcommand{\nn}{\nonumber}
\newcommand{\beq}{\begin{equation}}
\newcommand{\eeq}{\end{equation}}
\newcommand{\bea}{\begin{eqnarray}}
\newcommand{\eea}{\end{eqnarray}}

\newcommand{\1}{ \,  \raisebox{+0.14em}{{\hbox{{\rm \scriptsize ]}} \raisebox{-0.2em}{\kern-.8em\hbox{1}}}} \, }  

\title {A late times approximation for the SYK  spectral form factor}

\author{Matteo A. Cardella\footnote{matteo.cardella@unimi.it }\\
\normalsize\it Dipartimento di Fisica, Universit\`a degli Studi di Milano and INFN, \\
\normalsize \it  via Celoria 16, 20133 Milan, Italy \\
}

\begin{document}

\setcounter{tocdepth}{2}

\maketitle

\begin{abstract}
We find   a late times  approximation  for the SYK spectral form factor from  a   large $N$ steepest descent version  of the path integral  over two replica  collective  fields. Main ingredients are  a suitable uv regularization of the two replica kinetic operator, the property of its Fourier transform  and  some  spectral  analysis of the four point function two replica  ladder kernel.
\end{abstract}

\newpage

There is a tension  between the boundary holographic quantum mechanical description   of a region containing a black hole and its  bulk   semiclassical gravity description. While the former has a finite dimensional Hilbert space, whose discreteness   manifests
at times related to  coarse grain  scales  in the Hamiltonian spectrum, the bulk   has a continuous nature, in terms of classical fields.
In particular, this  tension manifests  in the behavior of  correlation functions at late times.
A boundary  thermal  correlator cannot decay to zero at late times, because of the finiteness  of the Hilbert space \cite{Dyson:2002pf},\cite{Goheer:2002vf},\cite{Barbon:2003aq}.
 On the other hand, this  same correlator computed by the bulk dictionary does go to zero at large separation times \cite{Horowitz:1999jd}.
A bulk  correlator  describes a certain number of  particles, scattered  by the
black hole horizon. The larger is the separation time between in and out states, the  larger  is the boost
for  particles  in reaching the horizon. The larger is their gravitational  boost, the longer is   the  tail of their  wave functions  beyond the horizon. As a consequence,  the scattering  reflection coefficient dies off in the infinite separation time limit. In other words, in  a semiclassical description,  the scattering amplitude for  particles by the black hole decays to zero  in the large separation time limit, in contradiction to  a   finite entropy  boundary quantum mechanics.
 This  is a manifestation  of the black hole information paradox \cite{Maldacena:2001kr}.
It is therefore of interest to consider systems where this tension can be quantitatively analyzed and hopefully resolved to some extent,  by comparing
 quantities in both sides of the correspondence.

\vspace{.3 cm}

A quantity that suits for this analysis \cite{Papadodimas:2015xma} is  the spectral form factor (SFF)  $Z(\beta - iT)Z(\beta + iT)$, an analytic continuation to time $T$ of the square of the imaginary time   thermal partition function at inverse temperature $\beta$. SFF is related to the two point correlation function  of pairs of  energy eigenvalues of the quantum system.
It gives information  in the  time variable $T$ on   correlations  among pairs of  eigenvalues at the  energy scale related to $T$.  For example, for a certain coarse  graining energy scale, a correlation dependence
 $\langle \rho(E) \rho(E') \rangle  \propto (E - E')^{-2}$ corresponds to a linearly growing in time $T$ form factor,
 at the related times scale.
 On the other hand,  a  $\langle \rho(E) \rho(E') \rangle \rightarrow 0$  behavior  at an  higher level of energy  coarse  graining (eigenvalues  level repulsion), corresponds to a constant value for the SFF at later times scales. 

\vspace{.3 cm}

 In the context of the  Sachdev-Ye-Kitaev model (SYK)  \cite{kitaev:talk1},\cite{kitaev:talk2},\cite{Sachdev-Ye93},\cite{Georges},\cite{Sachdev10},\cite{Sachdev10b},\cite{Sachdev}, numerical analysis  show \cite{Cotler2017}  a late times behavior  of the SYK spectral form factor\footnote{see also \cite{Garcia-Garcia:2016mno} for earlier related work. Recent works related to the SYK SFF include \cite{Winer:2020gdp},\cite{Khramtsov:2020bvs},\cite{delCampo:2017bzr}.} typical of certain  ensembles of Random Matrix Theory.  In particular, \cite{Cotler2017} have shown the existence of a linear ramp in $T$ rising behavior  that saturates at late times  to a constant plateau. Such a behavior is observed in various quantum chaotic  systems. In mathematics, there is the so called  Montgomery phenomenon, that involves the positions  of the  Riemann zeta function zeros on the critical line.  A repulsion among too closed zeros of the zeta function on the critical line is observed, that   is accurately reproduced by a correlator for  pairs of zeroes involving
 a Sinc function. It accounts  for the  ramp regime  and  the   plateau saturation shown by the Riemann zeta  spectral form factor.

\vspace{.3 cm}

 In \cite{Saad:2018bqo}   the SYK spectral form  factor is obtained  as a path integral over  two replica of collective continuous  bilocal fields with a complex coupling.   Various properties observed numerically  \cite{Cotler2017}, were deduced  \cite{Saad:2018bqo} by  a replica non-diagonal conformal  complex saddle. Since SYK is not exactly conformal even in the strongly coupled regime,
  the ramp part of the SYK  SFF  was  reproduced    \cite{Saad:2018bqo}  by employing  an  effective action in the two time reparametrization soft modes. This effective action is  given by   a complex linear combination of  two Schwarzians local actions. The two Schwarzians action  was introduced in the paper \cite{Saad:2018bqo} a bit out of the blue.   The authors  were based on  numerical evidence and  some general  indirect argument  \cite{DS}, which were not reported on their article.  In \cite{Cardella:2019kum},   the two Schwarzian  effective action was derived   by  direct computations on the two replica collective field  path integral. The method employed there,  relies on a  uv regularization of the two replica kinetic operator, inspired by a similar technique used  in regular SYK \cite{Kitaev:2017awl}.

\vspace{.3 cm}

  In this work, I develop further some of the techniques and computations presented in \cite{Cardella:2019kum}. In particular,  a late times $T$ approximation for the  SYK spectral form factor is obtained, by working on  a large $N$ approximation of the two replica path integral   \cite{Cardella:2019kum}. The large  $T$ approximation   is obtained here  by   spectral analysis of the four point function two replica ladder kernel. In particular,  the  explicit form of the ladder kernel   invariant eigenfunctions  is obtained. Then, an argument is provided, based on the Fourier transform of a uv regularization of the two replica  kinetic operator,  that   the ladder kernel   invariant eigenfunction  is
 the relevant quantity that governs  the  spectral form factor at large times.
Finally, by extracting   the  uv   behavior of this   eigenfunction,   the form   of the regularized source, needed for computing   the  large $T$ effective action is derived.
 The argument presented  at the end of this paper   provides an alternative route   w.r.t. one developed in \cite{Cardella:2019kum} for obtaining  the correct  form of the source that enters in the computation of the  late times SFF effective action.

\vspace{.9 cm}

In \cite{Cardella:2019kum}, we obtained  the following  large $N$ approximation for the SYK spectral form factor

\beq
Z(\beta + iT)Z(\beta - iT) \sim  e^{ \frac{N}{12 J^2} \langle s_{\alpha} |  \left( \delta_{\alpha \gamma} -  \tilde{K}_{\alpha \gamma }^{c}   \right)^{-1} \tilde{K}^{c}_{\gamma \beta}  | s_{\beta}\rangle  + \frac{N}{2} J_{\alpha}^{1/2} \langle s_{\alpha}  | |G_{\alpha}^{c}|^2 \rangle}, \label{Z2approx}
\eeq

which is valid in the strongly coupled nearly conformal regime.
Eq. (\ref{Z2approx}) was derived \cite{Cardella:2019kum}  by  steepest descent method on the  path integral for the spectral form factor, along contours  that go through the  replica non diagonal conformal saddle.
 The following  replica indexing \cite{Cardella:2019kum}  is used, $\alpha = 1, \dots, 4$, $\alpha = 1 = LL$, $\alpha = 2 = RR$, $\alpha = 3 = LR$, $\alpha = 4 = RL$.  The bilocal field  $s_{\alpha}(t,t')$  is a regularized version of the two replica kinetic operator $\delta_{ij}\delta(t - t')\partial_{t}$, $i,j= L,R$.  The   uv regularization replaces  the singular Dirac delta kernel $\delta(t - t')$  by   a smearing function $U(\xi(t,t'))$. It  implements the impossibility to resolve times intervals shorter then the SYK time scale $1 / J$ \cite{Kitaev:2017awl}. Moreover, the  derivative operator $\partial_t $ is substituted by a sum of contributions with different scalings \cite{Cardella:2019kum} that  are  analyzed  in terms of the spectrum of
the four point function two replica symmetrized ladder kernel at the conformal point
\beq
\tilde{K}_{\alpha \beta }^{c}  =  \tilde{K}^{c}_{ij,kl}(t_1, t_2 ; t_3, t_4) \equiv  3 J^2 J_{ij}^{1/2} J_{kl}^{1/2} |G_{ij}^{c}(t_{12})| G_{ik}^{c}(t_{13}) G_{jl}^{c}(t_{24}) |G_{kl}^{c}(t_{34})|, \qquad i,j = L,R. \label{Ksim}
\eeq

In the following,  we will  extract a late times $T$ approximation for (\ref{Z2approx}). To this aim,

Let us    expand the  source

\beq
s_{\alpha}(t,t') = \sum_{h} c_{h} \psi_{\alpha}^{h}(t,t') U( \xi_{\alpha}(t,t')). \label{sou}
\eeq

in terms of  eigenfunctions of  $\tilde{K}^{c}_{\alpha \beta}$

\beq
 \tilde{K}^{c}_{\alpha \beta}(t_1, t_2 ; t,t')\psi_{\beta}^{h}(t,t') =  \tilde{k}_{SFF}(h)\psi_{\alpha}^{h}(t_1 ,t_2).  \label{eighenfK}
\eeq

In (\ref{eighenfK}) $h$ labels  the eigenvalues of the  $SL_{diag}(2)$ Casimir, the diagonal subgroup of  $SL_{L}(2) \times S_{R}(2)$   that survives the  spontaneous breaking of the twofold time diffeomorphisms    by the conformal  replica non-diagonal saddle \cite{Saad:2018bqo}. Details and properties of  the smearing function $U( \xi_{\alpha}(t,t'))$  are given in \cite{Cardella:2019kum}.  For our purposes,   we find  convenient  to introduce  the  Fourier transform  $\tilde{U}(\eta)$

\beq
U(\xi) = \int \frac{d \eta}{ 2 \pi} e^{- i \xi \eta} \tilde{U}(\eta). \label{fourier}
\eeq

Since the smearing function $ U(\xi(t,t'))$ has support over  $\Delta \xi \sim O(T)$, its Fourier transform  $\tilde{U}(\eta)$
is non vanishing over an interval of length    $\Delta \eta  \sim  O\left( \frac{1}{T}\right)$, centered in $\eta = 0$.
  Therefore,  $\tilde{U}(\eta)$ becomes very narrow in the large $T$ limit.
We use (\ref{sou}) and  (\ref{fourier}) in  the first term at  the exponent of the spectral form factor (\ref{Z2approx})

\bea
&& \langle s_{\alpha} |  \left( \delta_{\alpha \gamma} -  \tilde{K}_{\alpha \gamma }^{c}   \right)^{-1} \tilde{K}^{c}_{\gamma \beta}  | s_{ \beta}\rangle    \nn \\
&=&  \sum_{h, h'} c_{h}c_{h'}  \int d t d t' (\psi_{\alpha}^{h'}(t,t'))^{*} U(\xi_{\alpha}(t,t')) \int \frac{d \eta}{ 2\pi}  \frac{\tilde{k}_{SFF}(h + i\eta )}{1 - \tilde{k}_{SFF}(h + i \eta)} \psi_{\alpha}^{h + i \eta }(t,t') \tilde U(\eta), \label{exponent3}
\eea

where \cite{Cardella:2019kum}
\beq
e^{- i \eta_{\alpha} \xi_{\alpha}}  =  \frac{1}{ \left(\zeta_{\alpha}(|t - t'|) \right)^{i\eta_{\alpha}}}.
\eeq

 In the  large $T$ limit,  one   expands   $\tilde{k}_{SFF}(h + i\eta)$   up to the first order in eta, 
$\tilde{k}_{SFF}(h + i\eta) = \tilde{k}_{SFF}(h) + i \eta \, \tilde{k}_{SFF}'(h)$. Then,  by evaluating the  $\eta$ integral   in (\ref{exponent3})  by    residues theorem, one finds

\beq
 \langle s_{\alpha} |  \left( \delta_{\alpha \gamma} -  \tilde{K}_{\alpha \gamma }^{c}   \right)^{-1} \tilde{K}^{c}_{\gamma \beta}  | s_{ \beta}\rangle
=  \sum_{h, h'} c_{h}c_{h'}  \int d t d t' (\psi_{\alpha}^{h'}(t,t'))^{*} U(\xi_{\alpha}(t,t')) \psi_{\alpha}^{h + i \eta_{h} }(t,t') \tilde U(\eta_{h}), \label{exponent}
\eeq

where

\beq
   \eta_h = i \frac{\tilde{k}_{SFF}(h) - 1}{\tilde{k}_{SFF}'(h)}. \label{etah}
\eeq

Since   $\tilde{U}(\eta)$ has support around $\eta = 0$ over an interval of length $O\left( \frac{1}{T} \right)$,  the only  non vanishing $\tilde{U}(\eta_h )$ in (\ref{exponent}) are those for

\beq
   |\eta_h|  = \left| \frac{\tilde{k}_{SFF}(h) - 1}{\tilde{k}_{SFF}'(h)} \right| <  \frac{\Delta \eta}{2} = O\left( \frac{1}{T}  \right) \label{etahmod}
\eeq

For  large enough time $T$, the only non vanishing term  in (\ref{exponent})
 reduces to  the  $\eta_{h} = 0$ one, which corresponds by eq. (\ref{etah})  to  $\tilde{k}_{SFF}(h_{*}) = 1$.
Therefore, we have the following large times $T$ approximation

\beq
 \langle s_{\alpha} |  \left( \delta_{\alpha \gamma} -  \tilde{K}_{\alpha \gamma }^{c}   \right)^{-1} \tilde{K}^{c}_{\gamma \beta}  | s_{ \beta}\rangle
\sim    \int_{0}^{T} d t \int_{0}^{T} d t' (\psi_{\alpha}^{h_{*}}(t,t'))^{*} U(\xi_{\alpha}(t,t')) \psi_{\alpha}^{h_{*} }(t,t'), \qquad T \rightarrow \infty. \label{exponent2}
\eeq

Therefore, we  have shown that in  the large $T$ limit, the only rescaled
smeared source which is effective at the exponent for the spectral form factor (\ref{Z2approx}) has  the following form

\beq
s_{\alpha}(t,t') = \psi_{\alpha}^{h_{*}}(t,t') U( \xi_{\alpha}(t,t')), \label{sou2}
\eeq

where $\psi_{\alpha}^{h_{*}}(t,t')$  is the invariant eigenfunction of the four points ladder symmetrized kernel (\ref{Ksim})

\beq
 \tilde{K}^{c}_{\alpha \beta}(t_1, t_2 ; t,t')\psi_{\beta}^{h_{*}}(t,t') =  \psi_{\alpha}^{h_{*}}(t_1 ,t_2).  \label{eighenfK1}
\eeq

Let us recall, (see \cite{Cardella:2019kum} for more  details), that the collective fields  action in  the spectral form factor  path integral splits as follows

\beq
Z(\beta + iT) Z(\beta -iT) = \int  \mathcal{D}G_{\alpha}  \mathcal{D}\Sigma_{\alpha}
 e^{N I_{c}(G_{\alpha}, \Sigma_{\alpha}) +  \int G_{\alpha}\sigma_{\alpha}}, \label{Z2split}
\eeq

where  $I_{c}$ is the time reparametrization invariant two replica critical action, \emph{without} the conformal symmetry breaking replica diagonal kinetic term  $\delta_{ij}\delta(t -t')\partial_t$.

On the other hand,

\beq
I_s =  \sum_{\alpha = 1,2} \int dt dt'  G_{\alpha}(t,t')\sigma_{\alpha}(t,t')   \label{Is}
\eeq

is the conformal  symmetric breaking source term,
where $\sigma_{\alpha}(t,t')$ is a regularized version of the  replica diagonal kinetic kernel $\delta_{ij} \delta(t - t')\partial_t$, which is related to the previously introduced $s_{\alpha}(t,t')$  by

\beq
\sigma_{\alpha}(t,t') =  |G_{\alpha}^{c}(t,t')| s_{\alpha}(t,t'),   \qquad   \alpha = 1,2. \label{sigmas}
\eeq

  The effective action in the time reparametrization soft modes $f_{L}(t)$, $f_{R}(t)$
 is computed from the non conformal  source action  (\ref{Is}).  In particular, it depends only on the  uv $|t - t'| \rightarrow 0$ behavior
 of the source  $\sigma_{\alpha}(t,t')$ . The reason for that  is that the failure of time reparametrization is a uv effect, as
  for   large times intervals, in the deep ir,  the  conformal saddle is a  very accurate approximation for  the actual Green function.
 As we have shown above,   in the large $T$ regime, the spectral form factor is dominated
  by a  source (\ref{sou2}) proportional to the  invariant eigenfunction $\psi_{\alpha}^{h_{*}}(t,t')$  (\ref{eighenfK1}) of the symmetrized  four point ladder kernel $\tilde{K}^{c}_{\alpha \beta}$.
 In the following  we  compute  explicitly
the diagonal components   $\alpha =1,2$ of $\psi_{\alpha}^{h_{*}}(t,t')$ and then extract their $| t - t'| \rightarrow 0$ behavior,
in order to obtain the form of the source that governs (\ref{Is}) in the large $T$ limit.
By the same method presented here,  one can also compute the off diagonals components $\alpha = 3,4$ of   $\psi_{\alpha}^{h_{*}}(t,t')$, we will not do it here, since only the diagonal components of the source matter  $\sigma_{\alpha}(t,t')$ in the source action (\ref{Is}).

Let us consider the    conformal Schwinger-Dyson equations for the spectral form factor \cite{Saad:2019lba},\cite{Cardella:2019kum}. An infinitesimal
variation of the bilocal fields by a  time reparametrization is  still a solution of the conformal saddle point equations,
due to conformal symmetry.  After some manipulation, one gets to  the following condition

\beq
(1 -  K^{c}_{\alpha \beta})  \delta_{\epsilon} G_{\beta}^{c} = 0,  \label{SDK}
\eeq

where  $K^{c}_{\alpha \beta}$ is the following  \emph{non symmetrized} version of the  four points ladder kernel

\beq
K^{c}_{ij,kl}(t_1, t_2 ; t_3, t_4) =  3 J^2  J_{kl}  G_{ik}^{c}(t_{13}) G_{jl}^{c}(t_{24}) (G_{kl}^{c}(t_{34}))^{2}  \label{K}
\eeq

and  $\delta_{\epsilon} G_{\alpha}^{c}$ is the first order variation of $G^{c}_{\alpha}(t,t')$ by  a linear diffeomorphism
$t \rightarrow   t  +  \epsilon(t)$. Eq. (\ref{SDK}) provides also  a way to compute the invariant eigenfunction of the symmetrized  kernel (\ref{Ksim}),
since  one  finds  that

\beq
 \psi_{\alpha}^{h_{*}}(t,t') =   J_{\alpha}^{1/2}|G_{\alpha}^{c}|\delta_{\epsilon}G^{c}_{\alpha}   \label{eigK}
\eeq

by (\ref{SDK})  is the invariant eigenfunction of  $\tilde{K}^{c}_{\alpha \beta}$.

We are now in the position of being able to compute the diagonal components $\alpha =1,2$ of  $\psi_{\alpha}^{h_{*}}(t,t')$ by using  eq. (\ref{eigK}).  A  linear time  reparametrization $t \rightarrow t + \epsilon(t)$ on  $G_{\alpha}(t,t')$, $\alpha = 1,2$
gives to the first order in $\epsilon(t)$

\beq
\delta_{\epsilon}G_{\alpha}(t,t')  =  \frac{1}{4} (\epsilon'(t) + \epsilon'(t'))G_{\alpha}(t,t') + (\epsilon(t) - \epsilon(t'))\partial_{t}G_{\alpha}(t,t') \label{varG}
\eeq

We use  the diagonal components of the spectral form factor conformal saddle \cite{Saad:2018bqo},\cite{Cardella:2019kum}

\beq
G_{\alpha}^{c}(t,t') \propto  \frac{sgn(t- t')}{\left|\frac{\tilde{\beta}_{aux}}{\pi} \sinh \left( \frac{\pi}{\tilde{\beta}_{aux}}(t -t')    \right)\right|^{1/2}},     \qquad   \alpha = 1,2,
\eeq

and its derivative

\beq
\partial_{t} G^{c}_{\alpha}(t,t') \propto - \frac{1}{2} \frac{sgn(t- t')}{\left|\frac{\tilde{\beta}_{aux}}{\pi} \sinh \left( \frac{\pi}{\tilde{\beta}_{aux}}(t -t')    \right)\right|^{3/2}}\cosh \left( \frac{\pi}{\tilde{\beta}_{aux}}(t -t')\right)  \qquad   \alpha = 1,2.
\eeq

We  express the linear time reparmetrization as a Fourier integral

\beq
\epsilon(t) =  \int \frac{d \omega}{2\pi} \, \,  e^{- i \omega t} \, \tilde{\epsilon}(\omega).  \label{epsilon}
\eeq

A  variation of the Green function by  a phase   $\epsilon(t) =  e^{- i \omega t} $,  by  eq. (\ref{varG})
gives

\beq
\delta_{\epsilon}G^{c}_{\alpha}  \propto  - i \frac{\omega}{4} ( e^{- i \omega t}  +  e^{- i \omega t'})G^{c}_{\alpha}(t,t') +  ( e^{- i \omega t}  -  e^{- i \omega t'})\partial_{t}G^{c}_{\alpha}(t,t'). \label{varG2}
\eeq

After  a little manipulation,  we get to  the following fixed $\omega$ form  for the  diagonal components of the invariant eigenfunction of the symmetrized ladder kernel

\bea
&& \psi_{\alpha}^{h_{*}, \omega}(t,t') =   J_{\alpha}^{1/2}|G_{\alpha}^{c}(t,t')| \delta_{\epsilon}G^{c}_{\alpha}  \nn \\
&=& i e^{-i \omega \frac{t + t'}{2}}\frac{sgn(t- t')}{\left|\frac{\tilde{\beta}_{aux}}{\pi} \sinh \left( \frac{\pi}{\tilde{\beta}_{aux}}(t -t')    \right)\right|} \left( \frac{\pi}{\tilde{\beta}_{aux}}  \frac{\sin\left( \omega \frac{t - t'}{2}   \right)}{  \tanh \left( \frac{\pi}{\tilde{\beta}_{aux}}(t - t') \right)} - \frac{\omega}{2} \cos\left( \omega \frac{t - t'}{2}\right)   \right), \nn \\
&& \alpha = 1,2. \label{eigfexpl}
\eea

On the other hand, by linearity (\ref{epsilon}),  the more general invariant  eigenfunction  of $\tilde{K}^{c}_{\alpha \beta}$ has the following   diagonal components

\beq
\psi_{\alpha}^{h_{*}}(t,t') = \int \frac{d \omega}{2 \pi} \, \tilde{\epsilon}(\omega) \,  \psi_{\alpha}^{h_{*}, \omega}(t,t'), \qquad  \alpha = 1,2, \label{eighf}
\eeq

with  $\tilde{\epsilon}(\omega)$ being the Fourier transform (\ref{epsilon}) of  $\epsilon(t)$.
Therfore, we have gotten   a continuous family of eigenfunctions       $\{  \psi_{\alpha}^{h_{*}, \omega}(t,t') \}_{\omega \in I_{\omega}}$ of the ladder kernel  with eigenvalue one,
labeled by $\omega \in I_{\omega} =  ( - \frac{\Delta_{\omega}}{2}, \frac{\Delta_{\omega}}{2})$, with $\Delta_{\omega} = O\left( \frac{1}{T}\right)$,
and (\ref{eighf}) is a generic linear superposition.

We want to extract the uv behavior  $|t - t'| \rightarrow 0$ in (\ref{eighf}), since by (\ref{sou2}) we know that it   enters in the computation of the large $T$ effective action for the spectral form factor through  the action term (\ref{Is}).
 Let us keep in mind that the source (\ref{sou2}) is uv regularized by the smeared function
$U(\xi(t,t'))$ that vanishes for $|t - t'| <  1/J$. On the other hand, $\tilde{\epsilon}(\omega)$ has support over an interval centered in $\omega = 0$ of length $|I_{\omega}| =  O\left( \frac{1}{T} \right)$.
In the limit $T \rightarrow \infty$ one can take  the following approximation

\beq
\psi_{\alpha}^{h_{*}}(t,t') = \int \frac{d \omega}{2 \pi} \, \tilde{\epsilon}(\omega) \,  \psi_{\alpha}^{h_{*}, \omega}(t,t'), \
\sim |I_{\omega}|\omega_{*} \tilde{\epsilon}(\omega_{*})\hat{\psi}_{\alpha}^{h_{*}}(t,t')  = -i \epsilon'(t_{\epsilon})\hat{\psi}_{\alpha}^{h_{*}}(t,t'), \label{estimate}
\eeq

where  $\omega_{*} \in I_{\omega}$, $t_{\epsilon} > 0$, and

\beq
\hat{\psi}_{\alpha}^{h_{*}}(t,t') = i \frac{sgn(t- t')}{\left|\frac{\tilde{\beta}_{aux}}{\pi} \sinh \left( \frac{\pi}{\tilde{\beta}_{aux}}(t -t')    \right)\right|} \left( \frac{\pi}{\tilde{\beta}_{aux}}  \frac{1}{  \tanh \left( \frac{\pi}{\tilde{\beta}_{aux}}(t - t') \right)} - \frac{1}{2}   \right).  \label{soueff}
\eeq

 Eq. (\ref{soueff}) is a valid approximation as far as $\tilde{\beta}_{aux} << T$. On the other hand,
$\epsilon'(t_{\epsilon})$ is regarded to be a renormalized quantity. Its meaning  follows from eq. (\ref{epsilon}) by the integral averaged theorem that was  used  in (\ref{estimate}).

By taking the uv $|t  - t'|\rightarrow 0 $  in   eq. (\ref{soueff}) we get   to the following form  for the replica diagonal components of  the effective  source in the large $T \rightarrow \infty$ limit

\beq
s_{\alpha}(t,t') \propto \frac{sgn(t- t')}{\left|\frac{\tilde{\beta}_{aux}}{\pi} \sinh \left( \frac{\pi}{\tilde{\beta}_{aux}}(t -t')    \right)\right|^{2}}U(\xi(t,t'))   \qquad   \alpha = 1,2.   \label{souUV}
\eeq

This is the quantity that enters in the computation of the effective action for the spectral form factor in the $T \rightarrow \infty$ limit through eq. (\ref{sigmas})  $\sigma_{\alpha}(t,t') =  |G_{\alpha}^{c}(t,t')|  s_{\alpha}(t,t')$
in the conformal symmetry breaking source term (\ref{Is}). Moreover, this same quantity governs the large $T$ approximation for the spectral form factor path integral (\ref{Z2approx}).
In \cite{Cardella:2019kum} the uv form of the smeared source displayed in eq. (\ref{souUV}), was obtained   by a different argument.
and used there  for computing   the two Schwarzians action in the time reparametrization soft modes  $f_{L}(t)$, $f_{R}(t)$.

\bibliography{SYK}
\bibliographystyle{JHEP}

\end{document}